\documentclass[9pt,twocolumn,twoside,a4paper]{article}
\usepackage{amsmath}
\usepackage{textcomp}
\usepackage{graphicx}

\title{Self-referenced characterization of space-time couplings\\ in near single-cycle laser pulses}

\author{T.~Witting, %
D.R.~Austin, %
T.~Barilot, %
D.~Greening, %
P.~Matia-Hernando,\\ %
D.~Walke, %
J.P.~Marangos, %
J.W.G.~Tisch}

\date{Compiled \today}

\begin{document}
\newlength{\figurewidth}
\setlength{\figurewidth}{0.97\linewidth}

\maketitle
	\textbf{We report on the characterization of space-time couplings in high energy sub-2-cycle 770\,nm laser pulses using a self-referencing single-shot method. Using spatially-encoded arrangement filter-based spectral phase interferometry for direct electric field reconstruction (SEA-F-SPIDER) we characterize few-cycle pulses with a wavefront rotation of $2.8\times 10^{11}$\,rev/sec (1.38\,mrad per half-cycle) and pulses with pulse front tilts ranging from to -0.33\,fs/\textmu{}m to -3.03\,fs/\textmu{}m.}

Near single cycle lasers are an essential tool for attosecond science and open new frontiers in non-linear optics~\cite{brabec_intense_2000,krausz_attosecond_2009}. It is usually assumed that the electric field of a laser pulse does not exhibit space-time coupling, i.e. the field can be written as the product of a spatially dependent field and a temporally dependent field $E(x,y,t) = E(x,y) E(t)$. Most characterization techniques measure the electric field of the beam at one spatial point $E(x_0,y_0,t)$ or average over a sub-sample of the beam selected by an aperture.
Some recently introduced attosecond pulse generation techniques such as the attosecond lighthouse for attosecond pulse gating in high harmonic generation (HHG) from plasma mirrors~\cite{wheeler_attosecond_2012,quere_applications_2014} and gases~\cite{kim_photonic_2013}, or non-collinear gating of HHG~\cite{heyl_noncollinear_2014,louisy_gating_2015} make use of intentional spatio-temporal distortions where $E(x,y,t) \neq E(x,y) E(t)$, but $E(x,y,t) = E(x,y,t+\xi_x x+\xi_y y)$ with the spatio-temporal coupling coefficients $\xi_x$ and $\xi_y$. Space-time couplings can also occur in few-cycle pulses produced in a hollow fibre (HCF) pulse compression system by angular misalignments or imperfect mode-match of the input beam to the fundamental fiber mode.

The ability to measure STC is important for the optimization of laser sources and experiments, and also for accurate modelling. Some spatio-temporal pulse characterization techniques exist, but most rely on interfering the STC pulse with a known reference pulse and spatially scanning a probe along $x$ and $y$, and are thus not single shot capable. Recent examples include SEA-TADPOLE, where a frequency resolved optical gating (FROG) measurement of a reference pulse is combined with spectral interferometry and an optical fibre scanned in $x$ and $y$ through the spatial profile of the unknown pulse~\cite{bowlan_crossed-beam_2006}. A variant is STARFISH which follows the same concept, but replacing FROG by the dispersion scan (DSCAN) technique to characterize the reference pulse~\cite{alonso_spatiotemporal_2012}.
The complete retrieval of the optical amplitude and phase using the ($k,\omega$) spectrum (CROAK) technique combines multiple frequency and angle resolved measurements of the pulse spectrum in the near and far fields with a Gerchberg-Saxton algorithm~\cite{bragheri_complete_2008}.
The shackled FROG technique combines a single FROG measurement with spatial amplitude and phase measurements via the Hartmann–Shack method~\cite{rubino_spatiotemporal_2009}.
DSCAN has been combined with point diffraction interferometry and Fourier transform spectrometry to characterize sub-8-fs pulses from an oscillator~\cite{miranda_spatiotemporal_2014}.
The only single frame method is STRIPED FISH~\cite{guang_complete_2014}. However, the beam is sampled at only a few spatially disjoint points and it is not a zero-additional phase measurement. A review of STC in ultrashort laser pulses and spatio-temporal pulse characterization methods can be found in a recent paper by Akturk et al~\cite{akturk_spatio-temporal_2010}.

A powerful technique for the characterization of ultrafast laser pulses is spectral phase interferometry for direct electric field reconstruction (SPIDER)~\cite{iaconis_spectral_1998}. Due to the direct algebraic pulse reconstruction and the fact that 1D data traces $S(x_0,y_0,\omega)$ describe a 1D pulse field $E(x_0,y_0,t)$, SPIDER lends itself to the extension to multiple spatial dimensions. SPIDER has been combined with additional spatial shearing to allow for the direct space–time characterization of ultrashort optical pulses in the 60\,fs range~\cite{dorrer_direct_2002,dorrer_spatio-temporal_2002}. In this letter we show  that SEA-F-SPIDER is capable of reconstructing space-time couplings in sub-2-cycle pulses without an additional spatial shearing interferometer. We measure for the first time ultrafast wavefront rotation and pulse front tilt of a sub-2-cycle pulse. The ability of SEA-F-SPIDER to reconstruct near-single-cycle pulses has been demonstrated previously~\cite{witting_characterization_2011}.
Reconstruction of pulse-front distortions (apart from a linear $x$-$t$-couplings) by SEA-F-SPIDER have been shown in~\cite{witting_spatio-temporal_2012}. Here we extend this to also include the linear (pulse-front tilt) by adding a correction term to the SPIDER reconstruction routine~\cite{wyatt_analysis_2009}. The frequency and thus the $k$-vector of ancilla B change from calibration to non-zero shear setting, which has to be taken into account by an extra term $\theta\Omega x/c$ in the description of the SEA-F-SPIDER interferogram:
\begin{eqnarray}
\label{eq:seaspidertrace}
S(x,\omega) &=& |E(x,\omega)|^2 +  |E(x,\omega-\Omega)|^2 +\\
\nonumber &&2|E(x,\omega)|\,|E(x,\omega-\Omega)| \times\\
\nonumber&&\cos{[\phi(x,\omega)-\phi(x,\omega-\Omega)+\Delta k x + \theta\Omega x/c]} \end{eqnarray}
The $\theta\Omega x/c$ term has to be removed for accurate reconstruction of the pulse-front-tilt. Knowledge of the exact frequencies 
$\omega_{\text{B},\Omega=0}$ and $\omega_{\text{B},\Omega\neq0}$ is essential, and is enabled by the direct spectral filtering approach of the ancilla preparation~\cite{witting_improved_2009}.

For the experiments in this paper we used a laser system capable of producing pulses with energies of up to 0.6\,mJ and durations down to 3.5\,fs at a repetition rate of 1\,kHz, produced by spectral broadening of 1\,mJ 30\,fs pulses from a Ti:Sapphire CPA system (Femtolasers) in a differentially pumped hollow fibre system filled with 3\,bar of neon on the laser exit side. Pulse compression is achieved by 10 reflections in a double-angle chirped mirror compressor (Ultrafast Innovations). A detailed description of the laser and pulse compression system can be found in~\cite{frank_invited_2012} and~\cite{okell_carrier-envelope_2013}.
\begin{figure}[!ht]
	\centering
	\includegraphics[width=\figurewidth]{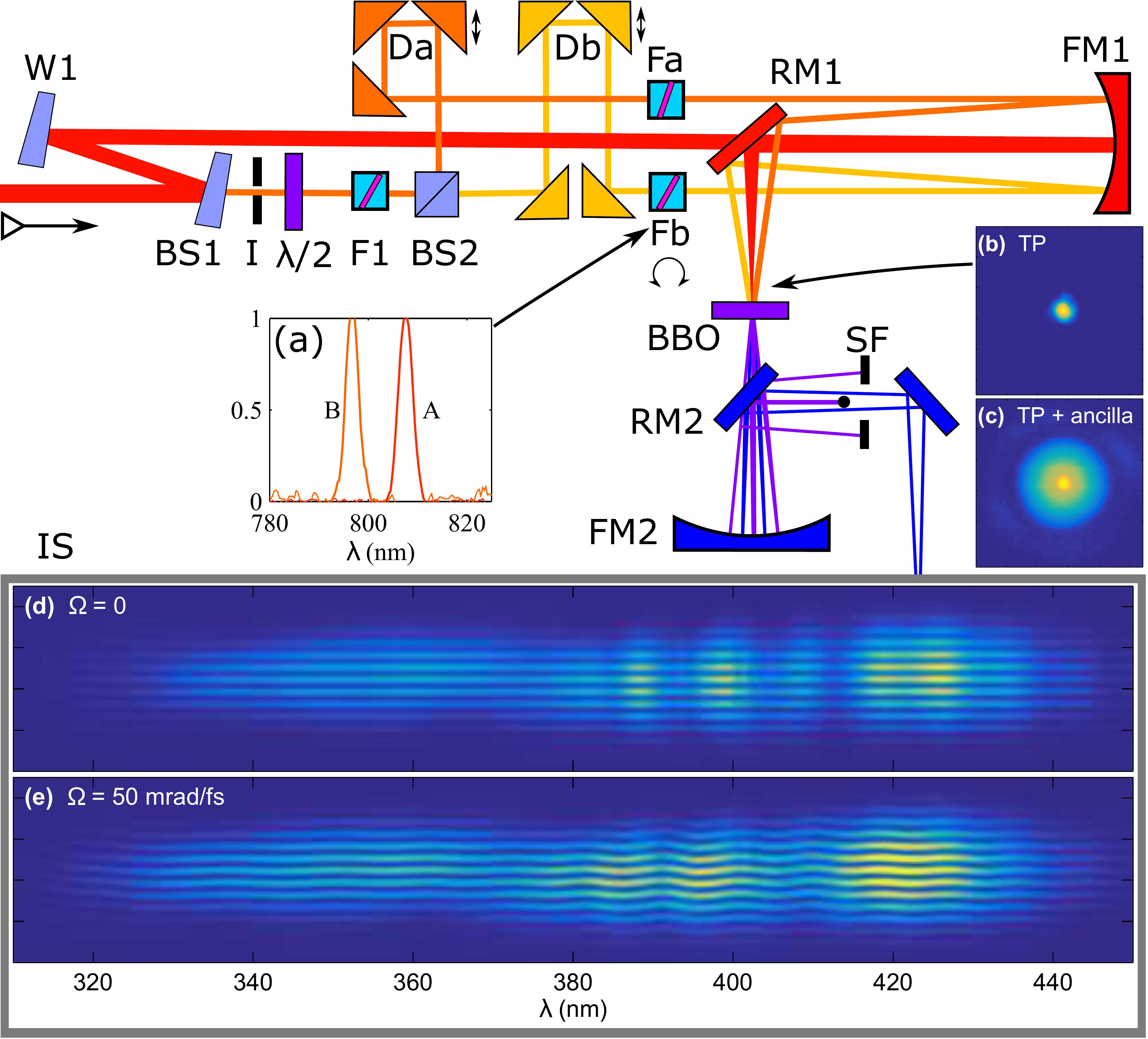}
	\caption{Experimental setup of the SEA-F-SPIDER device. Insets: ancilla spectra (a). focal images of TP (b) and ancilla (c). Calibration trace (d). SPIDER trace for a large shear (e). For a detailed description please refer to the text.}
	\label{fig:setup}
\end{figure}
Fig.\ref{fig:setup} shows the experimental setup of our SEA-F-SPIDER apparatus. The test pulse enters the device and a small fraction is picked off by an uncoated beam sampler BS1. Use of a second uncoated wedge W1 instead of a mirror ensures low enough intensity at the crystal to avoid spurious non-linear effects.
A half wave waveplate ($\lambda/2$) rotates the ancilla polarization for type-II upconversion in the crystal BBO ($\theta=43$\textdegree, L=20\textmu{}m). The ancilla beam is divided into two arms A and B with adjustable delay lines Da and Db to set the temporal overlap with the test pulse. The delays only have to be set once. Fa and Fb are narrowband bandpass filters (Semrock MaxLine LL01-808). Typical transmission spectra are shown in Fig.\ref{fig:setup}(a). To remove unwanted leakage of the filters Fa and Fb at wavelengths below 600\,nm we use long-pass filter F1 (Thorlabs, FGL665).
The rotation of Fa and Fb is motorised and we use fully automated data acquisition after initial calibration of transmission wavelength versus angle.  The filter rotation axis is parallel to the plane of beam separation to avoid a change in $\Delta k x$ with angle. The three beams, ancilla A, TP, and ancilla B are aligned parallel and focussed into the BBO crystal by focussing mirror FM1 ($f=300$mm). Focal spot images are shown in insets (b) and (c). Note the larger focal spot size of the ancilla due to aperturing by iris I. This ensured that the TP is upconverted with a uniform field. The two SPIDER sum-frequency signal beams are then re-imaged onto a home-built imaging spectrometer IS with stigmatic spatial and spectral imaging~\cite{austin_broadband_2009}. Spatial filter SF removes the beams of the fundamental and second harmonic of the TP and ancilla beams. The device measured the beam spatially in the horizontal direction.
We record a calibration trace at $\Omega=0$ as shown in Fig.\ref{fig:setup}(d). The fringes are flat and allow the extraction of $\Delta k x$. Fig.\ref{fig:setup}(e) shows a typical data trace. The fringe modulations following the gradient of the spectral phase are clearly visible for large shears (here $\Omega=50\,\text{mrad/fs}$).

\begin{figure}[!ht]
	\centering
	\includegraphics[width=\figurewidth]{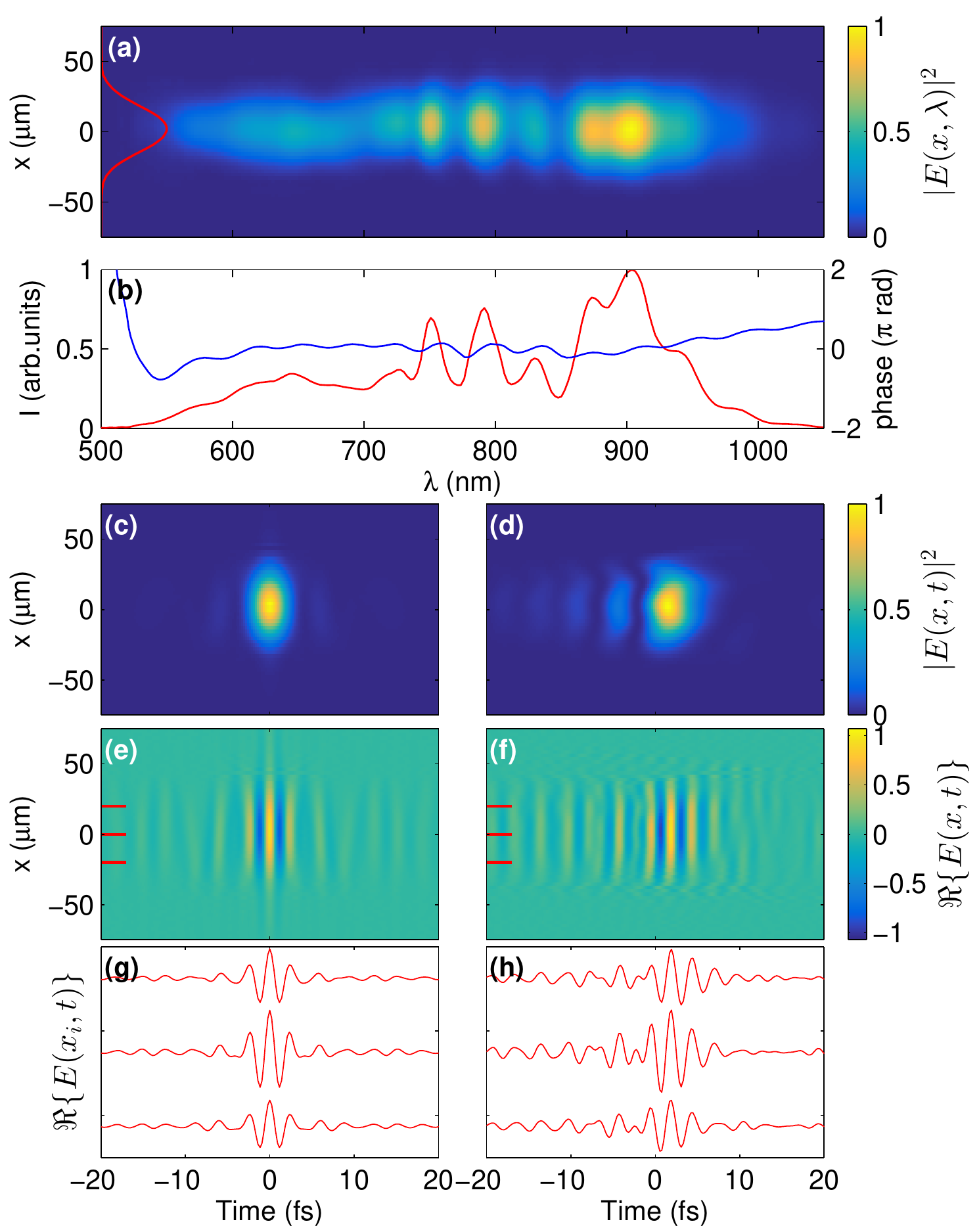}
	\caption{Spatio-temporal reconstruction of a sub-2-cycle pulse. Please refer to the text for details.}
	\label{fig:results_cleanpulse}
\end{figure}

Fig. \ref{fig:results_cleanpulse} shows a spatio-temporal measurement of a sub-2-cycle pulse. In (a) we show the intensity distribution in the spatio-spectral domain $|E(x,\omega)|^2$ on a linear color scale. The beam is virtually free from spatial chirp.
In (b) the spatially integrated 1D spectrum is shown as red line with the spectral phase at the beam center as blue line. The pulse has a cubic component of 5fs$^3$. At 540\,nm the phase increases strongly. This is the end of the design range of our chirped mirrors. At 850\,nm a distinct kink is visible where the phase changes slope, being reasonably linear below and above 850\,nm. We attribute this a manufacturing defect in production of the multilayer structure as this shows up in all our pulse measurements, independent of the configuration of our hollow fiber system and settings of the Dazzler pulse shaper.
Both the small cubic component and the phase kink contribute to	 the three small pre-pulses visible in the space-time intensity plot (d). The pulse duration is 3.8\,fs (1.6 optical cycles (OC)). The Fourier limited pulse intensity $|E(x,t)|^2$ has a duration of 3.3\,fs (1.4\,OC) and is shown in (c). The real part of the complex electric field $\Re\{E(x,t)\}$ is displayed in (e) (Fourier-limit) and (f). (g) and (h) show the electric field at three well-separated spatial positions $x_i$ of the beam. The pulse front tilt is negligible and the temporal profile varies very little across the spatial profile.

\begin{figure}[!ht]
	\centering
	\includegraphics[width=\figurewidth]{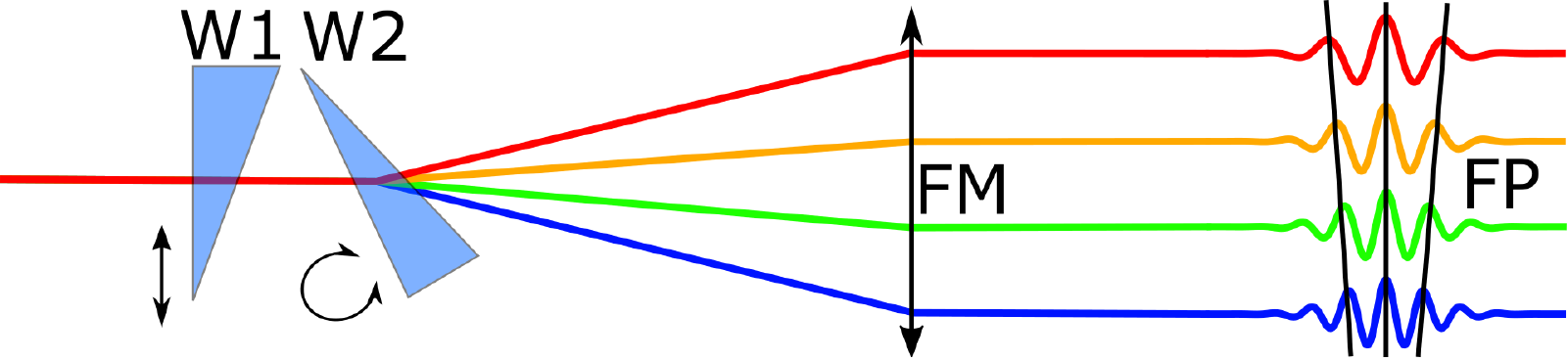}
	\caption{Generation of wavefront rotation by intentional misalignment of the dispersion fine tuning wedges. Please refer to the text for details.}
	\label{fig:WFR_sketch}
\end{figure}

To demonstrate the ability to recover space-time couplings we first turn to ultrafast wavefront rotation (WFR). This STC can be employed to enable spatial gating of attosecond pulses~\cite{wheeler_attosecond_2012,kim_photonic_2013}. Fig.\ref{fig:WFR_sketch} shows a schematic of our experimental setup. We intentionally misalign the two fused silica wedges used for pulse compression fine tuning. By tilting one wedge in the horizontal plane $x$ we introduce angular dispersion into the collimated beam. In the focal plane of a focussing mirror this is mapped to a spatially dependent spectrum (spatial chirp). It is easy to see that the spatial dependence of the center angular frequency $\omega_0=\omega_0(x)$ leads to a rotation of the pulse wavefronts. The rotating wavefront is indicated by thin black lines for three half-cycles.

\begin{figure}[!ht]
	\centering
	\includegraphics[width=\figurewidth]{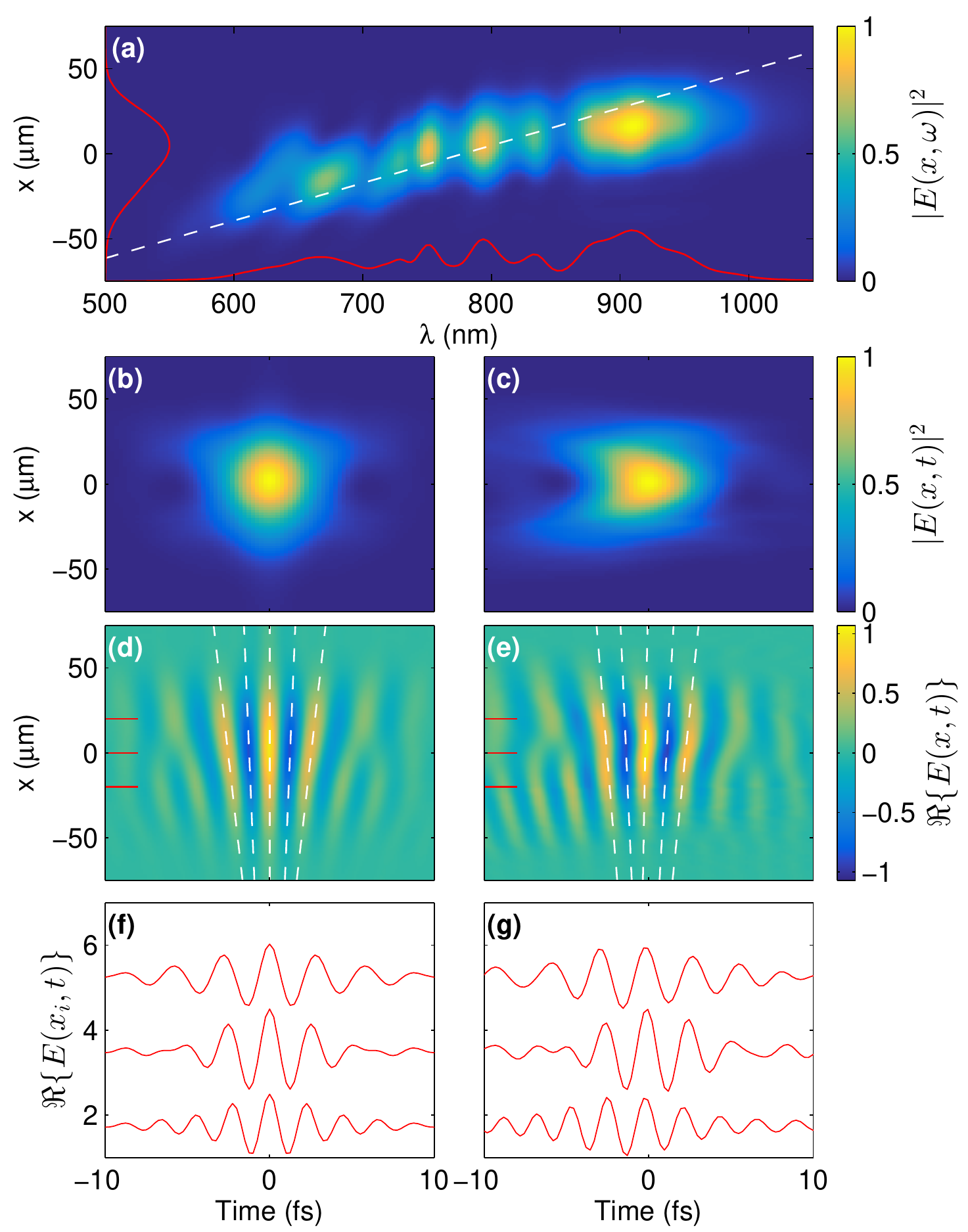}
	\caption{Few-cycle pulse with wavefront rotation. Please refer to the text for details.}
	\label{fig:WFR_results}
\end{figure}

Fig. \ref{fig:WFR_results} shows the SEA-F-SPIDER measurement of a pulse with WFR. In (a) the intensity in the spatio-spectral domain $|E(x,\omega)|^2$ is shown with linear color scale. The spatial chirp is clearly visible. The centre wavelength varies nearly linear with position. A few additional distortions are visible at 640\,nm. The spatially integrated spectrum matches the spectrum of the original pulse (Fig.\ref{fig:results_cleanpulse}).
The Fourier limited pulse intensity $|E(x,t)|^2$ is shown in (b) with the actual pulse in (c). Due to the spatial chirp rate of $d\omega/dx = -15.4$\,mrad/fs/\textmu{}m ($d\lambda/dx =  8.4$\,nm/\textmu{}m) the pulse duration is increased by 37\% from 3.8\,fs to 5.2\,fs (2.0\,OC). The Fourier limited duration of the WFR pulse is 4.5\,fs (1.7\,OC).
Note that the WFR pulse is also spatially enlarged compared to the STC free pulse. The original pulse has a focal spot size ($1/e^2$ intensity) of 61\textmu{}m. The spatially chirped WFR pulse has a 32\% larger spot size of 81\,\textmu{}m, 
which agrees well with the approximation $w^\prime = (1/w_0^2 - \xi^2/\Delta\omega^2)^{-1/2} = 82$\,\textmu{}m\,\cite{gu_spatial_2004}, with the initial spot size $w_0$, and spectral bandwidth $\Delta\omega$.
The electric field of the Fourier limited and the actual pulses are shown in (d) and (e) with clearly visible wavefront rotation. The WFR speed can be estimated from the spatial chirp rate by expressing the spatial chirp rate $d\omega_0/dx$ as a temporal field period change rate $dT_0/dx$ with $T_0=2\pi/\omega_0$. The WFR speed is then $dT_0/dx \cdot c / ( \langle T_0\rangle /2\pi)$. For the measured spatial chirp rate of -15.4\,mrad/fs/\textmu{}m we estimate a WFR speed of $2.8\times10^{11}$\,rev/sec, or perhaps more descriptive with respect to attosecond pulse generation, as wavefront angle rotation per half cycle of 1.38\,mrad/$T_{1/2}$ for a mean half-cycle time of $\langle T_{1/2} \rangle = $2.56\,fs ($\langle \omega_0 \rangle = $2.46\,rad/fs (766\,nm)).
The measured rotation speed of the Fourier limit (Fig.\ref{fig:WFR_results}(d)) is $2.2\times10^{11}$\,rev/sec, or 1.09\,mrad/$T_{1/2}$. The actual electric field (Fig.\ref{fig:WFR_results}(e)) has a WFR speed of $2.0\times10^{11}$\,rev/sec, or 0.98\,mrad/$T_{1/2}$.
Electric field lineouts at three different spatial positions $x_i$ are given in (f) and (g) showing the position dependent change in carrier frequency. 

\begin{figure}[ht]
	\centering
	\includegraphics[width=\figurewidth]{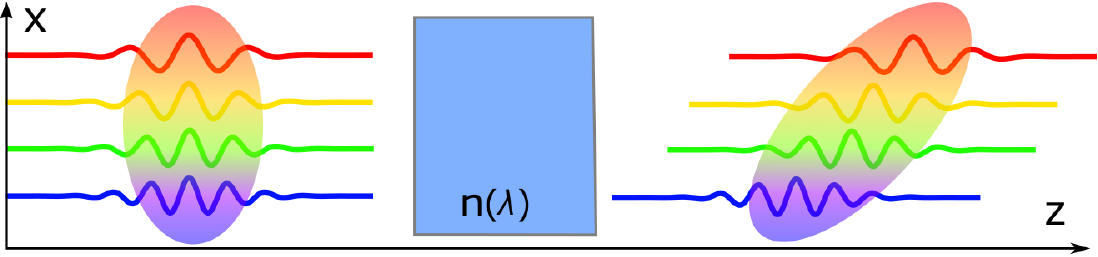}
	\caption{Generation of pulse front tilt by combination of spatial chirp and disperion.}
	\label{fig:PFT_sketch}
\end{figure}

A second important space time coupling is pulse front tilt (PFT). PFT can be written as a linear coupling of space and time $E(x,y,t) = E(x,y)E(t-px)$ with the pulse front tilt parametrized by $p$. To generate artificial and easily controllable PFT we combine a fixed spatial chirp with a varying amount of dispersion~\cite{akturk_pulse-front_2004}. Fig.\ref{fig:PFT_sketch} illustrates this effect.  

\begin{figure}[ht]
	\centering
	\includegraphics[width=\figurewidth]{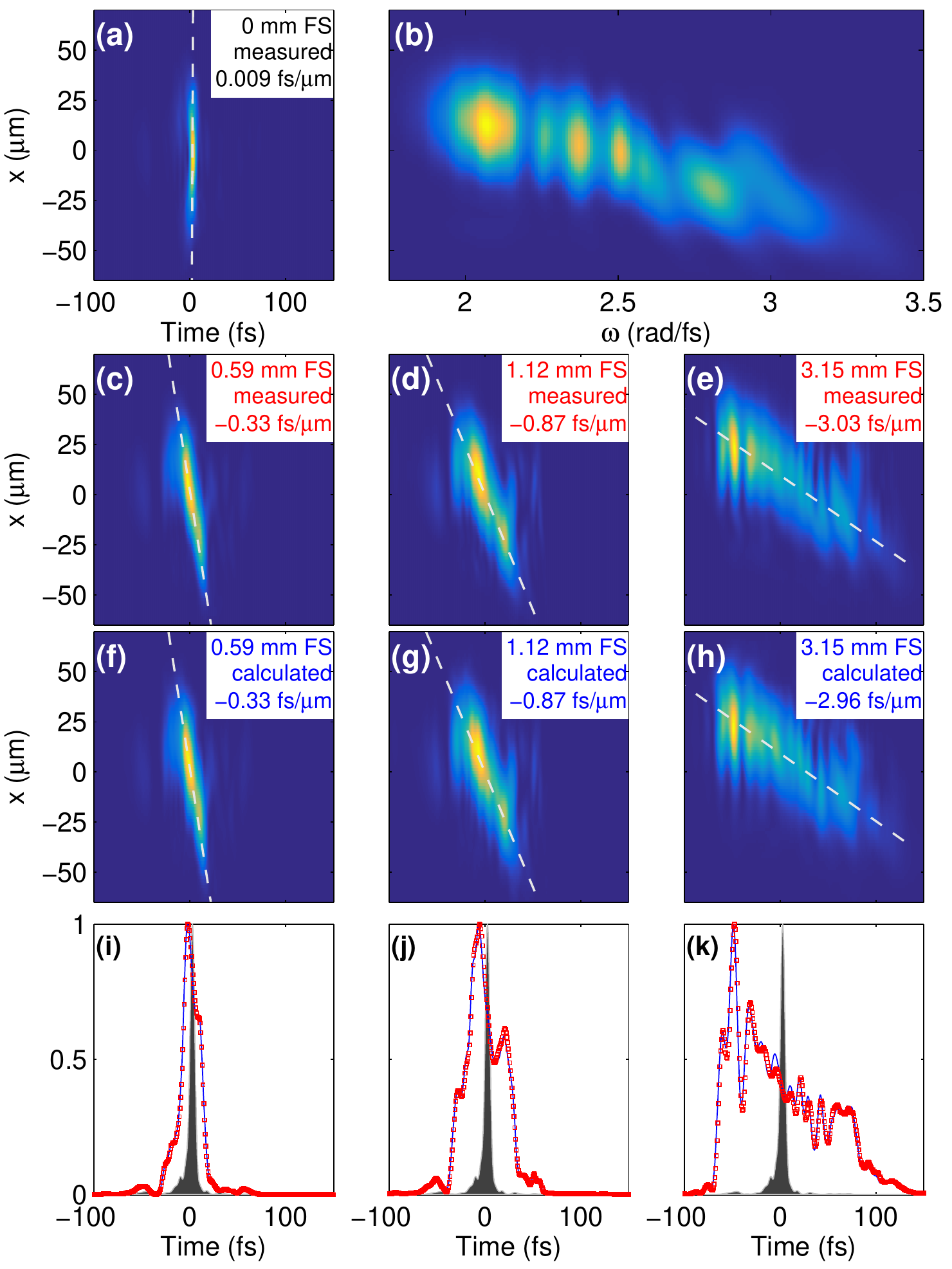}
	\caption{Few-cycle pulse with varying degrees of pulse front tilt. Details in text.}
	\label{fig:PFT_results}
\end{figure}

In Fig.\ref{fig:PFT_results} we show measurements of pulses with varying degrees of pulse front tilt. (a) and (b) show the initial condition: a spatially chirped pulse with optimum compression set by fine-tuning the wedge W1 (see Fig.\ref{fig:WFR_sketch}). Fig.\ref{fig:PFT_results}(a) shows the pulse intensity in space-time and (b) the spectral intensity  $|E(x,\omega)|^2$. The pulse front tilt is 0.009\,fs/\textmu{}m. The following three columns show pulses with PFT. Column 1 (c),(f),(i) correspond to the pulse of (a)/(b) with the added dispersion of 0.59\,mm of fused silica. The SEA-F-SPIDER reconstruction of $|E(x,t)|^2$ is displayed in (c). (f) shows a simulation using the known material dispersion for fused silica. A small temporal stretch and the PFT are clearly visible. The measured and calculated pulse front tilts are -0.329\,fs/\textmu{}m and -0.327\,fs/\textmu{}m. In (i) we show the spatially integrated temporal intensity profiles $\int|E(x,t)|^2dx$ for the compressed PFT free pulse (black area), the marginal of the measured pulse (red squares), and the marginal of the simulation (blue line). The FWHM pulse duration is 20.4\,fs (measurement), and 20.8\,fs (simulation).
The second column shows results for 1.12\,mm of added fused silica glass. The pulse front tilt is -0.869\,fs/\textmu{}m (-0.875\,fs/\textmu{}m simulation). The FWHM pulse duration is 43.4\,fs (44.3\,fs simulation).
The third column for 3.15\,mm of added fused silica glass. The pulse front tilt is now -3.035\,fs/\textmu{}m (-2.963\,fs/\textmu{}m simulation). The full width at $1/e^2$ pulse duration is 150.0\,fs (150.6\,fs simulation).
The agreement between measured PFT pulses and the calculations is excellent. Only for the very large chirp (e),(h),(k) the predicted pulse front is smaller by 2.4\,\% compared to the measurement. The disagreement between simulation and measurement is $<1\%$ for the two pulses with smaller PFT.

In conclusion, we have shown experimentally that SEA-F-SPIDER is capable of uncovering spatio-temporal distortions in sub-2-cycle laser pulses. We characterized sub-2-cycle pulses with minimal space time couplings, and also sub-2-cycle pulses with two versions of space-time couplings. We demonstrate for the first time spatio-temporal measurements of ultrafast wavefront rotation and pulse front tilt of a sub-2-cycle pulse including larger pulse front tilts caused by combination of spatial chirp and material dispersion. We envisage that our method will be useful for the optimization of advanced attosecond gating schemes such as the attosecond lighthouse, and also for the characterization of few-cycle laser systems where spatio-temporal distortions are expected (e.g. OPCPA) and the detection and removal of those is essential for improving their performance.

\section*{Funding Information}

Engineering and Physical Sciences Research Council (EPSRC), EP/I032517/1.


\begin{thebibliography}{10}
	\newcommand{\enquote}[1]{``#1''}
	
	\bibitem{brabec_intense_2000}
	T.~Brabec and F.~Krausz, Rev. Mod. Phys. \textbf{72}, 545 (2000).
	
	\bibitem{krausz_attosecond_2009}
	F.~Krausz and M.~Ivanov, Rev. Mod. Phys. \textbf{81}, 163 (2009).
	
	\bibitem{wheeler_attosecond_2012}
	J.~A. Wheeler, A.~Borot, S.~Monchocé, H.~Vincenti, A.~Ricci, A.~Malvache,
	R.~Lopez-Martens, and F.~Quéré, Nat Photon \textbf{6}, 829 (2012).
	
	\bibitem{quere_applications_2014}
	F.~Quéré, H.~Vincenti, A.~Borot, S.~Monchocé, T.~J. Hammond, K.~T. Kim,
	J.~A. Wheeler, C.~Zhang, T.~Ruchon, T.~Auguste, J.~F. Hergott, D.~M.
	Villeneuve, P.~B. Corkum, and R.~Lopez-Martens, J. Phys. B: At. Mol. Opt.
	Phys. \textbf{47}, 124004 (2014).
	
	\bibitem{kim_photonic_2013}
	K.~T. Kim, C.~Zhang, T.~Ruchon, J.-F. Hergott, T.~Auguste, D.~M. Villeneuve,
	P.~B. Corkum, and F.~Quéré, Nat Photon \textbf{7}, 651–656 (2013).
	
	\bibitem{heyl_noncollinear_2014}
	C.~M. Heyl, S.~N. Bengtsson, S.~Carlström, J.~Mauritsson, C.~L. Arnold, and
	A.~L'Huillier, New J. Phys. \textbf{16}, 052001 (2014).
	
	\bibitem{louisy_gating_2015}
	M.~Louisy, C.~L. Arnold, M.~Miranda, E.~W. Larsen, S.~N. Bengtsson, D.~Kroon,
	M.~Kotur, D.~Guénot, L.~Rading, P.~Rudawski, F.~Brizuela, F.~Campi, B.~Kim,
	A.~Jarnac, A.~Houard, J.~Mauritsson, P.~Johnsson, A.~L'Huillier, and C.~M.
	Heyl, Optica \textbf{2}, 563 (2015).
	
	\bibitem{bowlan_crossed-beam_2006}
	P.~Bowlan, P.~Gabolde, A.~Shreenath, K.~McGresham, R.~Trebino, and S.~Akturk,
	Opt. Express \textbf{14}, 11892 (2006).
	
	\bibitem{alonso_spatiotemporal_2012}
	B.~Alonso, M.~Miranda, i.~J. Sola, and H.~Crespo, Opt. Express \textbf{20},
	17880 (2012).
	
	\bibitem{bragheri_complete_2008}
	F.~Bragheri, D.~Faccio, F.~Bonaretti, A.~Lotti, M.~Clerici, O.~Jedrkiewicz,
	C.~Liberale, S.~Henin, L.~Tartara, V.~Degiorgio, and P.~Di~Trapani, Optics
	Letters \textbf{33}, 2952 (2008).
	
	\bibitem{rubino_spatiotemporal_2009}
	E.~Rubino, D.~Faccio, L.~Tartara, P.~K. Bates, O.~Chalus, M.~Clerici,
	F.~Bonaretti, J.~Biegert, and P.~Di~Trapani, Optics Letters \textbf{34}, 3854
	(2009).
	
	\bibitem{miranda_spatiotemporal_2014}
	M.~Miranda, M.~Kotur, P.~Rudawski, C.~Guo, A.~Harth, A.~L'Huillier, and C.~L.
	Arnold, Optics Letters \textbf{39}, 5142 (2014).
	
	\bibitem{guang_complete_2014}
	Z.~Guang, M.~Rhodes, M.~Davis, and R.~Trebino, Journal of the Optical Society
	of America B \textbf{31}, 2736 (2014).
	
	\bibitem{akturk_spatio-temporal_2010}
	S.~Akturk, X.~Gu, P.~Bowlan, and R.~Trebino, J. Opt. \textbf{12}, 093001
	(2010).
	
	\bibitem{iaconis_spectral_1998}
	C.~Iaconis and I.~A. Walmsley, Opt. Lett. \textbf{23}, 792 (1998).
	
	\bibitem{dorrer_direct_2002}
	C.~Dorrer, E.~M. Kosik, and I.~A. Walmsley, Opt. Lett. \textbf{27}, 548 (2002).
	
	\bibitem{dorrer_spatio-temporal_2002}
	C.~Dorrer, E.~Kosik, and I.~Walmsley, Applied Physics B: Lasers and Optics
	\textbf{74}, s209 (2002).
	
	\bibitem{witting_characterization_2011}
	T.~Witting, F.~Frank, C.~A. Arrell, W.~A. Okell, J.~P. Marangos, and J.~W.~G.
	Tisch, Optics Letters \textbf{36}, 1680 (2011).
	
	\bibitem{witting_spatio-temporal_2012}
	T.~Witting, S.~J. Weber, J.~W.~G. Tisch, and J.~P. Marangos, Opt. Express
	\textbf{20}, 27974 (2012).
	
	\bibitem{wyatt_analysis_2009}
	A.~Wyatt and I.~Walmsley, \enquote{Analysis of space-time coupling in
		{SEA}-{SPIDER} measurements,} in \enquote{European {Conference} on {Lasers}
		and {Electro}-{Optics} 2009. {CLEO} {Europe} 2009,}  (2009), pp. 1--1.
	
	\bibitem{witting_improved_2009}
	T.~Witting, D.~R. Austin, and I.~A. Walmsley, Opt. Lett. \textbf{34}, 881
	(2009).
	
	\bibitem{frank_invited_2012}
	F.~Frank, C.~Arrell, T.~Witting, W.~A. Okell, J.~McKenna, J.~S. Robinson, C.~A.
	Haworth, D.~Austin, H.~Teng, I.~A. Walmsley, J.~P. Marangos, and J.~W.~G.
	Tisch, Review of Scientific Instruments \textbf{83}, 071101 (2012).
	
	\bibitem{okell_carrier-envelope_2013}
	W.~A. Okell, T.~Witting, D.~Fabris, D.~Austin, M.~Bocoum, F.~Frank, A.~Ricci,
	A.~Jullien, D.~Walke, J.~P. Marangos, R.~Lopez-Martens, and J.~W.~G. Tisch,
	Opt. Lett. \textbf{38}, 3918 (2013).
	
	\bibitem{austin_broadband_2009}
	D.~R. Austin, T.~Witting, and I.~A. Walmsley, Appl. Opt. \textbf{48}, 3846
	(2009).
	
	\bibitem{gu_spatial_2004}
	X.~Gu, S.~Akturk, and R.~Trebino, Optics Communications \textbf{242}, 599
	(2004).
	
	\bibitem{akturk_pulse-front_2004}
	S.~Akturk, X.~Gu, E.~Zeek, and R.~Trebino, Optics Express \textbf{12}, 4399
	(2004).
	
\end{thebibliography}
\end{document}